\documentclass[notitlepage,reqno,a4paper]{article}
\usepackage{amscd}
\usepackage{mathrsfs}
\usepackage{stmaryrd}
\usepackage{amsmath}
\usepackage{amssymb}
\usepackage{graphicx}
\usepackage{verbatim}
\usepackage{indentfirst}
\usepackage{txfonts}
\usepackage{cite}

\begin{document}
\title{Applications of density matrix in the
fractional quantum mechanics}
\author{Jianping Dong\thanks{Email:dongjp.sdu@gmail.com} \\\\Department of Mathematics, College
of Science\\ Nanjing University of Aeronautics and Astronautics,
Nanjing 210016 \\P. R. China}
\date{}
\maketitle
\begin{abstract}
The many-body space fractional quantum system is studied using the
density matrix method. We give the new results of the Thomas-Fermi
model, and obtain the quantum pressure of the free electron gas.
 We also show the validity of the Hohenberg-Kohn theory in the space fractional quantum mechanics and generalize the
density functional theory to the fractional quantum mechanics.
\end{abstract}
\section{Introduction}
The famous Feynman path integral \cite{fey,klein} reveals the
relationship between fractals \cite{man} and quantum mechanics. The
background of the Feynman approach to quantum mechanics is a path
integral over Brownian paths. The Brownian paths are
non-differentiable, self-similar curves that have a fractal
dimension that is different from its topological dimension.
Recently, Laskin \cite{laskin1,laskin2,laskin3,laskin4,laskin5} used
L\'evy paths instead of the Brownian ones in the path integral and
obtained a space fractional Schr\"odinger equation (FSE) and then
developed fractional quantum mechanics. Following him, some other
generalizations to the standard Schr\"odinger equation appeared.
Naber \cite{naber} showed a time fractional Schr\"odinger equation ,
and then Wang and Xu \cite{wang} combined the two kinds of
fractional Schr\"odinger equations together to construct a
space-time fractional Schr\"odinger equation. Moreover, the
Fractional Heisenberg equation \cite{dong1,Rabei,tarasov} is also
developed.
 These generalizations may describe more extensive fractal phenomena in the microcosmic
world. \par There is a physical reason for the merger of fractional
calculus \cite{pod,kilbas} with quantum mechanics. The Feynman path
integral formulation of quantum mechanics is based on a path
integral over Brownian paths. In diffusion theory, this can also be
done to generate the standard diffusion equation; however, there are
examples of many phenomena that are only properly described when
non-Brownian paths are considered. When this is done, the resulting
diffusion equation has factional derivatives
\cite{west1,metz,Manuel}. Due to the strong similarity between the
Schr\"odinger equation and the standard diffusion equation one might
expect modifications to the Schr\"odinger equation generated by
considering non-Brownian paths in the path integral derivation. This
gives the time-fractional, space-fractional, and
space-time-fractional Schrodinger equation \cite{naber,wang,dong1}.
\par The space-fractional Schr\"odinger equation
\cite{laskin3} obtained by Laskin reads (in three dimensions)
\begin{equation} i\hslash\frac{\partial\psi(\mathrm{\textbf{r}},t)}{\partial
t}=H_\alpha \psi(\mathrm{\textbf{r}},t), \label{fse1}
\end{equation}
where $\psi(\mathrm{\textbf{r}},t)$ is the time-dependent wave
function , and $H_\alpha \text{ }(1<\alpha\leq 2)$ is the fractional
Hamiltonian operator given by
\begin{equation}
H_\alpha=-D_{\alpha}
(\hslash\nabla)^{\alpha}+V(\mathrm{\textbf{r}},t). \label{fse2}
\end{equation}
Here $D_{\alpha}$ with physical dimension
$[D_{\alpha}]=\text{[Energy]}^{1-\alpha}\times
\text{[Length]}^\alpha \times \text{[Time]}^{-\alpha}$ is dependent
on $\alpha$ [$D_{\alpha}=1/2m$ for $\alpha=2$, $m$ denotes the mass
of a particle] and $(\hslash\nabla)^{\alpha}$ is the quantum Riesz
fractional operator \cite{kilbas,laskin1} defined by
\begin{equation}
(\hslash\nabla)^{\alpha}\psi(\mathrm{\textbf{r}},t)=-\frac1{(2\pi\hslash)^3}\int\mathrm{d}^3\mathrm{\textbf{p}}\text{e}^
{i\mathrm{\textbf{p}}\cdot\mathrm{\textbf{r}}/\hslash}|\mathrm{\textbf{p}}|^\alpha\int\text{e}^
{-i\mathrm{\textbf{p}}\cdot\mathrm{\textbf{r}}/\hslash}\psi(\mathrm{\textbf{r}},t)\mathrm{d}^3\mathrm{\textbf{r}}
\label{riesz1}.
\end{equation}
\par  So far, the research on the fractional quantum mechanics is based on solving
 the FSE. In Ref.\cite{laskin3,dong2,dong3,guo}, the exact solutions to the FSE with free particle, Dirac
delta potential, linear potential, and some other simple potential
fields are given. However, because of the double complexity of the
fractional calculus and  quantum mechanics, it is hard to solve the
FSE directly for most cases,  especially for the many-body quantum
systems. In this paper, we use the density matrix method to study
the many-body fractional quantum system. We give the density matrix
description of  the fractional quantum system, study the
Thomas-Fermi model under the framework of the fractional quantum
mechanics, and generalize the density functional theory to the
fractional quantum mechanics.
\section{Density matrix description of the fractional quantum system}
\label{sec1} We are interested in systems of $N$ identical particles
(electrons, say) moving in a given external field and interacting
with each other with pair forces. In most cases, one is concerned
with atoms and molecules without time-dependent interactions, so we
may focus on the time-independent FSE \cite{dong2}. For an isolated
N-electron atomic or molecular system in the Born-Oppenheimer
non-relativistic approximation \cite{Robert,helmut}, the FSE is given by
\begin{equation}
\hat{H}_\alpha\psiup=E\psiup, \label{}
\end{equation}
where $E$ is the electronic energy,
$\psiup=\psiup(\textbf{x}_1,\textbf{x}_2,\cdots,\textbf{x}_N)$ is
the time-dependent wave function ( the coordinates $\textbf{x}_i$ of
electron $i$ comprise space coordinates $\textbf{r}_i$ and spin
coordinates $\textbf{s}_i$ ), and  $\hat{H}_\alpha$ is the
fractional Hamiltonian operator,
\begin{equation}
\hat{H}_\alpha=\hat{T}+\hat{V}_{ne}+\hat{V}_{ee}, \label{halpha}
\end{equation}
where $\hat{T}=\sum_{i=1}^{N}D_\alpha
(-\hslash^2\Delta_i)^{\alpha/2}$ is the kinetic energy operator, $
\hat{V}_{ne}=\sum_{i=1}^{N}\upsilon(\textbf{r}_i) $is the
electron-nucleus attraction energy operator, in which
$\upsilon(\textbf{r}_i)=-\sum_{\vartheta}\dfrac{Z_\vartheta
e^2}{r_{\vartheta i}}$ is the external potential acting on electron
$i$, the potential due to nuclei of charges $Z_\vartheta$, and
$\hat{V}_{ee}=\sum_{i<j}^{N}\dfrac{e^2}{r_{ij}} $ is the
electron-electron repulsion energy operator. The total energy $W$ of
this system is the electronic energy $E$ plus the nucleus-nucleus
repulsion energy, $
\hat{V}_{nn}=\sum_{\vartheta<\varrho}\dfrac{Z_\vartheta
Z_\varrho e^2}{R_{\vartheta\varrho}}. $ That is, $ W=E+\hat{V}_{nn}. $
\par Now we introduce the density matrix \cite{Robert,helmut} to describe this quantum
system. Considering the $N$ electrons system in a pure state
$\psiup_N(\textbf{x}_1\textbf{x}_2\cdots\textbf{x}_N)$, then we can
define the density matrix as
\begin{equation}
\gamma_N(\textbf{x}_1^\prime\textbf{x}_2^\prime\cdots\textbf{x}_N^\prime,\textbf{x}_1\textbf{x}_2\cdots\textbf{x}_N)
=\psiup_N(\textbf{x}_1^\prime\textbf{x}_2^\prime\cdots\textbf{x}_N^\prime)\psiup_N^*(\textbf{x}_1\textbf{x}_2\cdots\textbf{x}_N),
\end{equation} which can be viewed as the coordinate representation
of the density operator,
\begin{equation} \hat{\gamma}_N=|\psiup_N\rangle \langle \psiup_N|,
\end{equation}where the Dirac notation is used. Note that $\hat{\gamma}_N$ is a projection operator, and
for normalized $\psiup_N$, \begin{equation}
\text{tr}(\hat{\gamma}_N)=\int\psiup_N(\textbf{x}^N)\psiup_N^*(\textbf{x}^N)d\textbf{x}^N=1,
\end{equation} in which $\text{tr}(\hat{\gamma}_N)$, the
trace of the operator $\hat{\gamma}_N$, is defined as the sum of
diagonal elements of the matrix representing $\hat{\gamma}_N$, or
the integral if the representation is continuous. In the state
$\psiup_N$, the expectation values of any observable $A$ can be
calculated by \cite{Robert}
\begin{equation}
\langle\hat{A}\rangle=\text{tr}(\hat{A}\hat{\gamma}_N)=\text{tr}(\hat{\gamma}_N\hat{A}),\label{avvalue}
\end{equation}where $\hat{A}$ is the Hermitian linear operator for
the observable $A$. In view of Eq.~(\ref{avvalue}), the density
operator $\hat{\gamma}_N$ carries the same information as the
N-electron wave function $|\psiup\rangle$. $\hat{\gamma}_N$ is an
operator in the same space as the vector. Note that while
$|\psiup_N\rangle$ is defined only up to an arbitrary phase factor,
$\hat{\gamma}_N$ for a state is unique. $\hat{\gamma}_N$ also is
Hermitian. Further more, we can define the reduced density matrices
for fermion systems \cite{Robert},
\begin{equation}
\begin{split}\gamma_p&(\textbf{x}_1^\prime\textbf{x}_2^\prime\cdots\textbf{x}_N^\prime,\textbf{x}_1\textbf{x}_2\cdots\textbf{x}_N)
=\begin{pmatrix}
  N \\
  p \\
\end{pmatrix}
\int\cdots \int\gamma_N(\textbf{x}_1^\prime\textbf{x}_2^\prime\cdots\textbf{x}_p^\prime
\\&\textbf{x}_{p+1}\cdots\textbf{x}_{N},\textbf{x}_1\textbf{x}_2
\cdots\textbf{x}_{p}\cdots\textbf{x}_N)d\textbf{x}_{p+1}\cdots
d\textbf{x}_{N},
\end{split} \end{equation}
where $\begin{pmatrix}
  N \\
  p \\
\end{pmatrix}$ is a binomial coefficient. In particular,
\begin{equation}
\gamma_2(\textbf{x}_1^\prime\textbf{x}_2^\prime,\textbf{x}_1\textbf{x}_2)
=\frac{N(N-1)}{2}
\int\cdots\int\gamma_N(\textbf{x}_1^\prime\textbf{x}_2^\prime\textbf{x}_3\cdots
\textbf{x}_{N},\textbf{x}_1\textbf{x}_2
\textbf{x}_{3}\cdots\textbf{x}_N)d\textbf{x}_{3}\cdots
d\textbf{x}_{N},
\end{equation} and
\begin{equation}
\gamma_1(\textbf{x}_1^\prime,\textbf{x}_1)
=N\int\cdots\int\gamma_N(\textbf{x}_1^\prime\textbf{x}_2\cdots\textbf{x}_{N},\textbf{x}_1\textbf{x}_2\cdots\textbf{x}_N)d\textbf{x}_{2}\cdots
d\textbf{x}_{N}.
\end{equation} Many operators of interest do not involve spin coordinates, for instance the Hamiltonian operators for atoms or molecules. This makes desirable
further reduction of the density matrices of the first and the
second orders. We define the first-order and second-order spinless
density matrices by
\begin{equation} \rho_1(\textbf{r}_1^\prime,\textbf{r}_1)
=\int\gamma_1(\textbf{r}_1^\prime\textbf{s}_1,\textbf{r}_1\textbf{s}_1)d\textbf{s}_{1}.
\end{equation} and
\begin{equation} \rho_2(\textbf{r}_1^\prime\textbf{r}_2^\prime,\textbf{r}_1\textbf{r}_2)
=\iint\gamma_2(\textbf{r}_1^\prime\textbf{s}_1\textbf{r}_2^\prime\textbf{s}_2,\textbf{r}_1\textbf{s}_1\textbf{r}_2\textbf{s}_2)
d\textbf{s}_{1}d\textbf{s}_{2}.
\end{equation}We also introduce a shorthand notation for the diagonal element of $\rho_1$ and $\rho_2$:
\begin{equation} \rho(\textbf{r}_1)=\rho_1(\textbf{r}_1,\textbf{r}_1),\rho_2(\textbf{r}_1,\textbf{r}_2)=\rho_2(\textbf{r}_1\textbf{r}_2,\textbf{r}_1\textbf{r}_2).
\end{equation}Note that the diagonal element of $\rho_1$ is just the electron density, because that
\begin{equation} \rho(\textbf{r}_1)=\rho_1(\textbf{r}_1,\textbf{r}_1)=N\int\cdots\int|\psiup|^2d\textbf{s}_{1}d\textbf{x}_{2}\cdots d\textbf{x}_{N}.
\end{equation} A useful relationship between the first and the second order spinless density matrices is given as follows,
\begin{equation} \rho_1(\textbf{r}_1^\prime,\textbf{r}_1)=\frac{2}{N-1}\int \rho_2(\textbf{r}_1^\prime\textbf{r}_2,\textbf{r}_1\textbf{r}_2)d\textbf{r}_{2}.
\end{equation}In particular,
\begin{equation} \rho(\textbf{r}_1)=\frac{2}{N-1}\int\rho_2(\textbf{r}_1,\textbf{r}_2)d\textbf{r}_{2}.
\end{equation} After these definitions, the expectation value, for an antisymmetric N-body wave function, of a local one-electron operator $\hat{O}_1=
\sum_{i=1}^{N}O_1(r_i)$, can be expressed as
\begin{equation} \langle\hat{O}_1\rangle=\text{tr}(\hat{O}_1\hat{\gamma}_N)=\int[O_1(\textbf{r}_1)
\rho_1(\textbf{r}_1^\prime,\textbf{r}_1)]_{\textbf{r}_1^\prime=\textbf{r}_1}d\textbf{r}_{1}.
\end{equation}Similarly, the expectation value of a two-electron operator $\hat{O}_2=\sum_{i<j}^{N}O_2(r_i,r_j)$ is
\begin{equation}
 \langle\hat{O}_2\rangle=\text{tr}(\hat{O}_2\hat{\gamma}_N)
=\int[O_2(\textbf{r}_1\textbf{r}_2)
\rho_2(\textbf{r}_1^\prime\textbf{r}_2,\textbf{r}_1\textbf{r}_2)]_{\textbf{r}_1^\prime=\textbf{r}_1,\textbf{r}_2^\prime=\textbf{r}_2}d\textbf{r}_{1}d\textbf{r}_{2}.
\end{equation}
Using the above results, combining all the parts, the expectation
value of the Hamiltonian (\ref{halpha}), is obtained,
\begin{equation}
\begin{split}
 E=&\text{tr}(\hat{H}_\alpha\hat{\gamma}_N)=E(\rho_1(\textbf{r}_1^\prime,\textbf{r}_1),\rho_2(\textbf{r}_1,\textbf{r}_2))
\\=&\int[D_\alpha
(-\hslash^2\Delta_{\textbf{r}})^{\alpha/2}
\rho_1(\textbf{r}^\prime,\textbf{r})]_{\textbf{r}^\prime=\textbf{r}}d\textbf{r}+\int
v(\textbf{r})\rho(\textbf{r})d\textbf{r}+\iint\frac{e^2}{r_{12}}
\rho_2(\textbf{r}_1,\textbf{r}_2)d\textbf{r}_{1}d\textbf{r}_{2}.\label{totalenergy}
\end{split}
\end{equation}The three terms in this formula represent respectively
the electronic kinetic energy, the nuclear-electron potential
energy, and the electron-electron potential energy. If we write
\begin{equation} \rho_2(\textbf{r}_1,\textbf{r}_2)=\frac12\rho(\textbf{r}_1)\rho(\textbf{r}_2)[1+h(\textbf{r}_1,\textbf{r}_2)].
\end{equation}Then the
third term in Eq.~(\ref{totalenergy}) can be rewritten as
\begin{equation}
 \iint\frac{e^2}{r_{12}}
\rho_2(\textbf{r}_1,\textbf{r}_2)d\textbf{r}_{2}d\textbf{r}_{1}=J[\rho]+\frac12\iint\frac{e^2}{r_{12}}\rho(\textbf{r}_1)
\rho_{xc}(\textbf{r}_1,\textbf{r}_2)d\textbf{r}_{1}d\textbf{r}_{2},
\end{equation}where
\begin{equation} J[\rho]=\frac12\iint\frac{e^2}{r_{12}}\rho(\textbf{r}_1)
\rho(\textbf{r}_2)d\textbf{r}_{1}d\textbf{r}_{2},
\end{equation} and
$\rho_{xc}(\textbf{r}_1,\textbf{r}_2)=\rho(\textbf{r}_2)h(\textbf{r}_1,\textbf{r}_2)$.
Here, $\rho_{xc}(\textbf{r}_1,\textbf{r}_2)$ is called the
exchange-correlation hole \cite{Robert,helmut}, and
\begin{equation}\int\rho_{xc}(\textbf{r}_1,\textbf{r}_2)d\textbf{r}_{2}=-1.
\end{equation}So far, the density matrix description of the
fractional quantum mechanics has been made. Now we begin to study
the fractional quantum mechanics by use of these results.
\section{The Thomas-Fermi model}
\label{sec2}In the standard quantum mechanics, the first attempt to
study the density functional theory of electron structure is the
works of Thomas and Fermi in the 1920s. Here, under the framework of the fractional quantum mechanics,
 we also start from the Thomas-Fermi Model \cite{Robert,helmut,Landau}. We divide
the space into many small cells, each of side $l$ and volume $\Delta
V=l^3$, each containing some fixed number of electrons $\Delta N$,
and we assume that the electron in each cell behave like independent
fermions at the temperature $0$K, with the cells independent of one
another. Then we first study the energy levels of a particle in each
cell. The particle is like in a three-dimensional infinite well,
\begin{equation} V(x,y,z)=\left\{
\begin{aligned} 0, &\quad \text{if }x,y,z\in(0,l);\\\infty,
&\quad\text{otherwise}.
\end{aligned} \right.
\label{v1}
\end{equation}
So we have the following FSE,
\begin{equation} D_\alpha
(-\hslash^2\Delta)^{\alpha/2}\psiup(\textbf{r})=E\psiup(\textbf{r}),\label{fseforfree}
\end{equation}with the boundary conditions
\begin{equation}
\psiup(\textbf{r})=0, \text{ when } x=l, \text{ or } y=l, \text{ or
} z=l, \text{ or } x=y=z=0.\label{boundarycon}
\end{equation}It is easy to prove that $\psiup(\textbf{r})=C
e^{i\textbf{k}\cdot\textbf{r}}$ is a basic solution to
Eq.~(\ref{fseforfree}). Here, $\textbf{k}=(k_x.,k_y,k_z)$,
satisfying
\begin{equation}
E=D_\alpha\hslash^\alpha|\textbf{k}|^\alpha.\label{energywithk}
\end{equation} The general solution to Eq.~(\ref{fseforfree}) can be
expressed by
\begin{equation}
\psiup(\textbf{r})=A\sin(\textbf{k}\cdot\textbf{r})+B\cos(\textbf{k}\cdot\textbf{r}).
\end{equation}
Considering the boundary condition (\ref{boundarycon}), we have
\begin{equation}
B=0, \sin(k_xL)=\sin(k_yL)=\sin(k_zL)=0,
\end{equation}which means
\begin{equation}
k_xL=n_x\pi, k_yL=n_y\pi,k_zL=n_z\pi,
\end{equation} where $n_x,n_y,n_z$ can be any integer. So $\textbf{k}=(k_x,k_y,k_z)=\dfrac{\pi}{L}(n_x,n_y,n_z)$. Then from
Eq.~(\ref{energywithk}), we get
\begin{equation}
E(n_x,n_y,n_z)=D_\alpha\hslash^\alpha|\textbf{k}|^\alpha=D_\alpha(\frac{\pi\hbar}{L})^\alpha
R^\alpha, \label{energy1}
\end{equation} where $R=\sqrt{n_x^2+n_y^2+n_z^2}.$ Considering the spin states of electrons, every space state $(n_x,n_y,n_z)$ can be occupied by
two electrons. For high quantum numbers, that is, for large $R$, the
number of distinct energy levels with energy smaller than
$E(n_x,n_y,n_z)$ can be approximated by the volume of one octant of
a sphere with radius $R$ in the space $(n_x,n_y,n_z)$. That's
because that every array of positive integers $(n_x,n_y,n_z)$
corresponds to one point in the first quadrant of the space
$(n_x,n_y,n_z)$ and there is only one grip point in per volume unit.
Therefore, the number of distinct quantum states, in the interval
$(R,R+dR)$, is given by
\begin{equation}
dN=2\cdot\frac18\cdot4\pi R^2dR=\pi R^2dR,
\end{equation}where the factor $2$ enters because each energy level is doubly occupied,by one electron with spin $\alpha$ and another with spin
$\beta$. By the Pauli exclusion principle, we know that the number
of distinct quantum states is also the number of the electrons. When
$N$ is large, the change of $N$ can be viewed as continuous.
Differentiating Eq.~(\ref{energy1}) yields
\begin{equation}
dE=\alpha D_\alpha(\frac{\pi\hbar}{L})^\alpha R^{\alpha-1}dR.
\label{}
\end{equation}Thus
\begin{equation}
dN=\pi R^2(\alpha D_\alpha)^{-1}(\frac{L}{\pi\hbar})^\alpha
R^{1-\alpha}dE, \label{dnde}
\end{equation}That is
\begin{equation}
\frac{dN}{dE}=\pi R^2(\alpha
D_\alpha)^{-1}(\frac{L}{\pi\hbar})^\alpha
R^{1-\alpha},\label{numberinede}
\end{equation} the right side of which gives the number of electrons with energy in the interval $(E,E+dE)$.  Thus,
 the number of electrons with energy smaller than the Fermi energy $E_f$ is
\begin{equation}
N=\int_{0}^{E_f}\frac{dN}{dE}dE=\frac\pi3(D_\alpha)^{-3/\alpha}(\frac{L}{\pi\hbar})^3(E_f)^{3/\alpha}.\label{number1}
\end{equation}
The total energy for the volume cell can now be calculated as
\begin{equation}
E=\int_{0}^{E_f}E\frac{dN}{dE}dE=\frac{3}{\alpha+3}(D_\alpha)^{-3/\alpha}(\frac{L}{\pi\hbar})^3(E_f)^{(\alpha+3)/\alpha},\label{energy2}
\end{equation}
which can be rewritten as
\begin{equation}
E=\frac{3}{\alpha+3}NE_f=\frac{3}{\alpha+3}D_\alpha\hbar^\alpha(3\pi^2)^{\alpha/3}(\frac{N}{L^3})^{\alpha/3+1}L^3.\label{energy3}
\end{equation} When the volume cell $V\rightarrow0$, $L^3=dV=d\textbf{r}$ and electronic density
$\rho=\dfrac{N}{L^3}=\dfrac{N}{V}=\rho(\textbf{r})$. From
Eq.~(\ref{energy3}), we obtain the total energy $dE$ for the volume
cell $d\textbf{r}$ as
\begin{equation}
dE=\frac{3}{\alpha+3}D_\alpha\hbar^\alpha(3\pi^2)^{\alpha/3}(\rho(\textbf{r}))^{\alpha/3+1}d\textbf{r}.\label{energy4}
\end{equation}
Adding the contributions from all cells, we find the total kinetic
energy to be
\begin{equation}
T_{TF}[\rho]=C_F\int(\rho(\textbf{r}))^{\alpha/3+1}d\textbf{r},\label{kineticenergy}
\end{equation} where $C_F=\dfrac{3}{\alpha+3}D_\alpha\hbar^\alpha(3\pi^2)^{\alpha/3}$. This is the kinetic energy functional for the Thomas-Fermi model in the
fractional quantum mechanics. When $\alpha=2$, by use of the atom
unit, we have
$T_{TF}[\rho]=C_F\int(\rho(\textbf{r}))^{5/3}d\textbf{r}$, and
$C_F=\frac{3}{10}(3\pi^2)^{2/3}\approx2.871,$
 which accords with the standard quantum mechanics \cite{Robert}.
\par Recalling Eq.~(\ref{totalenergy}), we can get an energy formula for an
atom in terms of electron density alone,
\begin{equation}
E_{TF}[\rho(\textbf{r})]=C_F\int(\rho(\textbf{r}))^{\alpha/3+1}d\textbf{r}-Ze^2\int\frac{\rho(\textbf{r})}{r}d\textbf{r}+\iint\frac{e^2}{r_{12}}
\rho_2(\textbf{r}_1,\textbf{r}_2)d\textbf{r}_{1}d\textbf{r}_{2}.\label{thomasenergy}
\end{equation}For the ground state of an atom of interest the electron density minimizes the energy functional $E_{TF}[\rho(\textbf{r})]$, under the constraint
 \begin{equation}
 N=N[\rho(\textbf{r})]=\int\rho(\textbf{r})d\textbf{r},\label{constraintforn}
 \end{equation} where N is the total number of electrons in the atom. By the method of Lagrange multipliers, the ground-state electron density must satisfy the
variational principle
\begin{equation}
\delta\left\{
E_{TF}[\rho(\textbf{r})]-\mu_{TF}\left(\int\rho(\textbf{r})d\textbf{r}-N\right)\right\}=0,
\end{equation} where $\mu_{TF}$ is the Lagrange multiplier.  After executing the variation, we can get
\begin{equation}
\mu_{TF}=\frac{\alpha+3}{\alpha}C_F\rho^{\alpha/3}(\textbf{r})-\varphi(\textbf{r}),\label{mutf}
\end{equation} where $\varphi(\textbf{r})$ is the electrostatic potential at point $\textbf{r}$ due to the nucleus and the entire electron distribution:
\begin{equation}
\varphi(\textbf{r})=\frac{Ze^2}{r}-\int\frac{e^2\rho(\textbf{r}_2)}{|\textbf{r}-\textbf{r}_2|}d\textbf{r}_{2}.\label{varphieq}
\end{equation}
Eq.~(\ref{mutf}) can be solved in conjunction with the constraint
(\ref{constraintforn}), and the resulting electron density then
inserted in (\ref{thomasenergy}) to give the total energy.  There are the results of the
Thomas-Fermi model for the atom in the fractional quantum mechanics.
Just as in the standard quantum mechanics, it is a simple model,
without considering the exchange and correlation energy terms. But
in all the results, the parameter $\alpha$ is contained. Proper
choose of $\alpha$ may provide better results than the standard
quantum mechanics.
\par
\begin{figure}[h]
 \includegraphics[width=8cm]{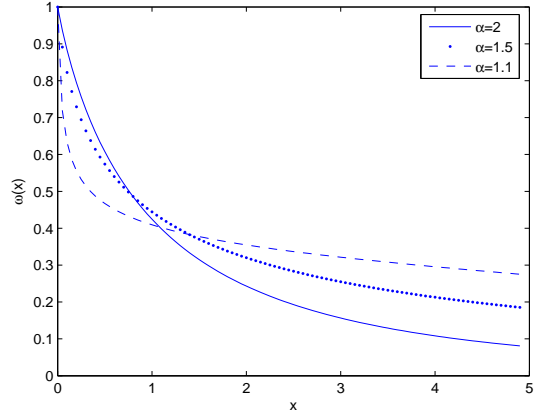}
 \centering
 \caption{The curves for the numerical solutions of $\omega(x)$ for
 $\alpha=1.1,1.5,2$.  $\alpha=2$ corresponds to the standard quantum mechanics (see Ref.~\cite{Robert,helmut,Landau}).
 }
 \label{fig1}
\end{figure}

 Now we turn to solve the equations (\ref{mutf}) and (\ref{varphieq}). For neutral atoms, when
$r\rightarrow\infty$, $\varphi(\textbf{r})\rightarrow0$ and
$\rho(\textbf{r})\rightarrow0$, so from Eq.~(\ref{mutf}), we have
$\mu_{TF}=0$, which means
\begin{equation}
\rho(\textbf{r})=C^{-3/\alpha}\varphi^{3/\alpha}(\textbf{r}),\label{densitywithvarphi}
\end{equation} where $C=C_F(\alpha+3)/\alpha$.
Then using the results of the Poisson equation \cite{Arfken}, the expression of
$\varphi(\textbf{r})$ in Eq.~(\ref{varphieq}) satisfies the
following equation
\begin{equation}
\nabla^2\varphi(\textbf{r})=4\pi e^2\rho(\textbf{r})-4\pi
e^2Z\delta(\textbf{r}),\label{poissonatom}
\end{equation} Substituting Eq.~(\ref{densitywithvarphi}) for $\rho(\textbf{r})$ gives
\begin{equation}
\nabla^2\varphi(\textbf{r})=4\pi
e^2C^{-3/\alpha}\varphi^{3/\alpha}(\textbf{r})-4\pi
e^2Z\delta(\textbf{r}).\label{subsub1}
\end{equation} Additionally, by the spherically symmetric of the atom, we
can write
\begin{equation}\varphi(\textbf{r})=\varphi(r)=\frac{Ze^2}{r}\omega(r).\label{varphiwithomega}
\end{equation}
Then we have
\begin{equation}
\nabla^2\varphi(\textbf{r})=\frac{Ze^2}{r}\frac{d^2\omega(r)}{dr^2}-4\pi
e^2Z\omega(0)\delta(\textbf{r}).\label{subsub2}
\end{equation}
Eq.~(\ref{subsub1}) minus Eq.~(\ref{subsub2}) side by side gives
\begin{equation}
\frac{d^2\omega(r)}{dr^2}=Mr^{1-\frac3\alpha}[\omega(r)]^{\frac3\alpha},
\end{equation} where $$M=4\pi C^{-3/\alpha}Z^{3/\alpha-1}e^{6/\alpha},$$
and the boundary condition $\omega(0)=1$ is introduced. Letting
$x=ar$  and $a=M^{\alpha/[3(\alpha-1)]}$, we obtain a differential equation for $\omega(x)$,
\begin{equation}
\frac{d^2\omega(x)}{dx^2}=x^{1-\frac3\alpha}[\omega(x)]^{\frac3\alpha},\label{omegax}
\end{equation}
with the boundary conditions $\omega(0)=1$ , $\omega(\infty)=0$.
This problem can be solved numerically. The numerical result is
given in Fig.~\ref{fig1} for different $\alpha$.
 From the figure, we know that for all $\alpha$, the curves decrease monotonously, and for a smaller $\alpha$,
the curve decreases more rapidly at the beginning, but more slowly
as $x\rightarrow\infty$. Note that Eq.~(\ref{omegax}) has an exact
solution $\omega(x)=Bx^A$, where $A=3(\alpha-1)/(\alpha-3)$, and
$B=[A(A-1)]^{\alpha/(3-\alpha)}$. This solution does not satisfy the
boundary condition at $x=0$, but can be a good approximate
expression for $\omega(x)$ when $x$ is large.

\section{Quantum pressure of the free electron gas}
By use of the results in the previous section, we can study the quantum
pressure of the electron gas in the fractional quantum mechanics.
Defining the variation of the volume of the electron gas by
$d\Omega$, the needed energy from outside by $dA$, then the pressure
$p$, of the electron gas, satisfies
\begin{equation} d\Omega=-pdA,
\end{equation} Meantime, the internal thermal energy increases by $dU=dA$, so
\begin{equation} p=-\frac{dU}{d\Omega}.\label{pressure1}
\end{equation}
For degenerate Fermi electron gas, using Eq.~(\ref{energy2}), we
have
\begin{equation} U=N\overline{E}=\frac{3N}{\alpha+3}E_f.
\end{equation} Thus
\begin{equation} dU=\frac{3N}{\alpha+3}dE_f.\label{inenergy}
\end{equation} Remembering Eq.~(\ref{number1}), we can get
\begin{equation}
E_f^{\frac3\alpha}=N\frac3\pi(D_\alpha)^{\frac3\alpha}\frac{(\pi\hbar)^3}{\Omega},\label{efomega}
\end{equation}where the volume of the electron gas $\Omega=L^3$. The
above equation can be converted into
\begin{equation} \frac3\alpha\ln E_f=constant-\ln{\Omega}.
\end{equation} Differentiating the both sides
gives
\begin{equation} d\Omega=-\frac3\alpha\frac{\Omega}{E_f}dE_f.\label{gasvolume}
\end{equation}
Then, with the help of Eqs.~(\ref{inenergy}) and (\ref{gasvolume}),
the pressure $p$ of the electron gas can be calculated by
Eq.~(\ref{pressure1}),
\begin{equation} p=-\frac{dU}{d\Omega}=\frac{\alpha}{\alpha+3}\frac{N}{\Omega}E_f=\frac{\alpha}{\alpha+3}\rho
E_f,\label{presure22}
\end{equation}where $\rho=N/\Omega$ denotes the density of the electron gas. Finally, combing  Eq.~(\ref{efomega}) and (\ref{presure22}),
 the pressure $p$  is expressed by the density $\rho$ as follows,
\begin{equation}
p=\frac{\alpha}{\alpha+3}(3\pi^2)^{\frac\alpha3}D_\alpha\hbar^\alpha\rho^{\frac\alpha3+1}.\label{finalp}
\end{equation}
This means there is a stabilizing internal pressure in the solid
object. But this pressure derives ultimately from the
anti-symmetrization requirement for the wave functions of identical
fermions, so it can be called the degeneracy pressure \cite{grif}.
When $\alpha=2$, Eq.~(\ref{finalp}) reduces to the result in the
standard quantum mechanics  \cite{grif}.

\section{Accurate theory: Hobenberg-Kohn theorems}
\label{sec3} In the previous sections, we study the many-body
fractional quantum system under the Thomas-Fermi model, an
exquisitely simple model. To get the accurate results, we recognize
the density functional theory \cite{Robert,helmut} just as in the standard quantum
mechanics. From the discussion in Ref.\cite{laskin3}, we know that
the hamiltonian $H_\alpha$ for the space fractional quantum system
is still a hermite operator. Employing the minimum-energy principle
for the ground state, we can easily prove that the Hohenberg-Kohn
theorems \cite{Robert,helmut} still hold in the space fractional quantum mechanics. We
now give the two theorems without proof, and the proof is exactly
the same with the standard quantum mechanics:
\\1.The electron density $\rho(\textbf{r})$ determines the ground-state wave
function and all the other electronic properties of the system.
\\2.For a trial density $\tilde{\rho}(\textbf{r})$, such that $\tilde{\rho}(\textbf{r})\geq0$ and
$\int\tilde{\rho}(\textbf{r})d\textbf{r}=N$,
\begin{equation}
E_0\leq E_\upsilon[\tilde{\rho}],
\end{equation}where $E_\upsilon[\tilde{\rho}]$ is the energy functional
of the quantum system with external potential
$\upsilon(\textbf{r})$. From Eq.~(\ref{totalenergy})in
Sec.~\ref{sec1}, the total energy can be written as
\begin{equation}
E_\upsilon[\rho]=T[\rho]+V_{ne}[\rho]+V_{ee}[\rho]=\int
v(\textbf{r})\rho(\textbf{r})d\textbf{r}+F_{HK}[\rho],
\end{equation} where $F_{HK}[\rho]=T[\rho]+V_{ee}[\rho]$.We may write $V_{ee}[\rho]=J[\rho]+\text{nonclassical term}$, where $J[\rho]$ is the classical repulsion. The nonclassical term is a very elusive, very important quantity; it is the major part of the "exchange-correlation energy".
Assuming differentiability of $E_\upsilon[\rho]$, the ground-state
electron density must satisfy the variational principle
\begin{equation}
\delta\left\{
E_{\upsilon}[\rho(\textbf{r})]-\mu\left(\int\rho(\textbf{r})d\textbf{r}-N\right)\right\}=0,\label{hkvariation}
\end{equation}which gives the Euler-Lagrange equation
\begin{equation}
\mu=\frac{\delta E_{\upsilon}}{\delta
\rho(\textbf{r})}=\upsilon(\textbf{r})+\frac{\delta
F_{HK}[\rho]}{\delta \rho(\textbf{r})}.
\end{equation} The quantity $\mu$ is the Lagrange multiplier and called the chemical potential. If we knew the exact $F_{HK}[\rho]$,
 Eq.~(\ref{hkvariation}) would be an exact equation for the ground-state electron density. As in the standard quantum mechanics,
  accurate calculational implementations of the density functional theory are far from easy to achieve, because of the unfortunate
fact that the functional $F_{HK}[\rho]$ is hard to come by in explicit form.
  To implement density functional theory, the Kohn-sham method \cite{Robert,helmut} can also be used here and the resulting Kohn-sham equation will contain a fractional operator.
Further research on this subject will be done in our future work.
\section{Conclusions}
\label{sec4} In this paper, the many-body space fractional quantum
system, governing by fractional Schr\"odinger equation obtained by
Laskin, is studied using the density matrix method. We give the
results of the Thomas-Fermi model, calculate the quantum pressure of
electron gas in the fractional quantum mechanics. We also show the
validity of the Hohenberg-Kohn theorems, and generalize the density
functional theory (DFT) to fractional  quantum mechanics. These
results provide the possibility to study the fractional quantum
system without solving the FSE directly.
\par The density functional theory is a
popular tool applied to study the electron structure of solids
\cite{Martin,Walsh,Michael}, but because the exchange-correlation
term is unknown and hard to define, there are always modifications
needed. The fractional quantum mechanics, which is built in the
sense of L\'evy flight, is a generlisation to the standard one,
 which is built in the sense of Brownian motion. The L\'evy flight is a natural generalization of the Brownian motion, and the fractal dimension of
Brownian paths is two, while that of L\'evy paths is $\alpha\text{ }(1<\alpha\leq2)$. The many-body quantum systems can exhibit the characteristic of fractal
 with the fractal dimension not equal to two, so it is reasonable to believe that the fractional Schr\"odinger equation can better describe the behavior
 of the microcosmic particles.
  The density functional theory in the fractional quantum mechanics may be an alternative way to study the structure of the many-body quantum systems,
and without making more modifications to the exchange-correlation terms, by choosing a proper parameter $\alpha$, the DFT in the fractional quantum mechanics
may provides better results than in the standard quantum mechanics, which needs our further studies.


\end{document}